# Temperature Effects on Brain Tissue in Compression

Badar Rashid[a], Michel Destrade[b,a], Michael Gilchrist[a*]

[a]School of Mechanical and Materials Engineering, University College Dublin, Belfield, Dublin 4, Ireland

[b]School of Mathematics, Statistics and Applied Mathematics, National University of Ireland Galway, Galway, Ireland

*Corresponding Author

Tel: + 353 1 716 1884/1991, + 353 91 49 2344  Fax: + 353 1 283 0534

Email: Badar.Rashid@ucdconnect.ie (B. Rashid), michael.gilchrist@ucd.ie (M.D. Gilchrist), michel.destrade@nuigalway.ie (M. Destrade)

**Abstract**

Extensive research has been carried out for at least 50 years to understand the mechanical properties of brain tissue in order to understand the mechanisms of traumatic brain injury (TBI). The observed large variability in experimental results may be due to the inhomogeneous nature of brain tissue and to the broad range of test conditions. However, test temperature is also considered as one of the factors influencing the properties of brain tissue. In this research, the mechanical properties of porcine brain have been investigated at $22^o$ C (room temperature) and at $37^o$ C (body temperature) while maintaining a constant preservation temperature of approximately $4 - 5^o$ C. Unconfined compression tests were performed at dynamic strain rates of 30 and 50/s using a custom made test apparatus. There was no significant difference ($p = 0.8559 \sim 0.9290$) between the average engineering stresses of the brain tissue at the two different temperature conditions. The results of this study should help to understand the behavior of brain tissue at different temperature conditions, particularly in unconfined compression tests.

# 1    Introduction

The human head has been identified as the most sensitive region frequently involved in life-threatening injuries such as road traffic accidents, sports accidents and falls. Intracranial brain deformation caused by rapid angular acceleration or blunt impact to the head during injurious events is responsible for traumatic brain injuries (TBIs), which are a leading cause of death or disability. Diffuse axonal injury (DAI) is a type of TBI which is characterized by microscopic damage to axons throughout the white matter of the brain, and focal lesions in the corpus callosum and rostral brainstem. To gain a better understanding of the mechanisms of TBI, several research groups have developed numerical models which contain detailed geometric descriptions of the anatomical features of the human head, in order to simulate and investigate internal dynamic responses to multiple loading conditions (Ho and Kleiven, 2009; Horgan and Gilchrist, 2003; Kleiven, 2007; Ruan et al., 1994; Zhang et al., 2001). However, the fidelity of these models is highly dependent on the accuracy of the material properties used to model biological tissues; therefore, a systematic research on the constitutive behavior of brain tissue under impact is essential.

Extensive research is still underway to understand the biomechanics of TBI. Several studies have been conducted to determine the range of strain and strain rates associated with DAI. DAI in animals and human has been hypothesized to occur at macroscopic shear strains of 10% – 50% and strain rates of approximately 10 – 50/s (Margulies et al., 1990; Meaney and Thibault, 1990) although locally the strains and strain rates could be much higher than their macroscopic values, due to the complex geometry and material inhomogeneities of brain tissue. The threshold strains predicted for injury range from 0.13 to 0.34 (Bain and Meaney, 2000). Similarly, tests have also been performed by applying strains within the range of 20%–70% and strain rates within the range of 20 – 90/s to create mild to severe axonal injuries (Pfister et al., 2003). Studies conducted by Morrison et al., (2006; 2003; 2000), suggested that brain cells are significantly damaged at strains > 0.10 and strain rates > 10/s.

Several research groups investigated the mechanical properties of brain tissue in order to establish constitutive relationships over a wide range of loading conditions. Mostly dynamic oscillatory shear tests were conducted over a frequency range of 0.1 to 10000 Hz (Arbogast et al., 1997; Bilston et al., 2001; Brands et al., 2004; Darvish and Crandall, 2001; Fallenstein et al., 1969; Ho and Kleiven, 2009; Hrapko et al., 2006; Nicolle et al., 2004; 2005; Ning et al., 2006; Peters et al., 1997; Prange and Margulies, 2002; Shen et al., 2006; Shuck and Advani, 1972; Takhounts et al., 1999; Thibault and Margulies, 1998) and unconfined compression tests (Cheng and Bilston, 2007; Estes and McElhaney, 1970; Franceschini et al., 2006; Miller and Chinzei, 1997; Pervin and Chen, 2009; Prange and Margulies, 2002; Tamura et al., 2007) while a limited number of tensile tests (Franceschini et al., 2006; Miller and Chinzei, 2002; Tamura et al., 2008; Velardi et al., 2006) were performed and again, the reported properties vary from study to study. The recorded variation in test results is probably related to the anisotropic and inhomogeneous nature of brain tissue, different test protocols and to the broad range of test conditions.

The effect of variable temperatures in the test protocols adopted by various research groups is an important issue, although only limited studies have been reported (Brands et al., 2000; Peters et al., 1997; Shen et al., 2006). Hrapko et al., (2008) used an eccentric rotational rheometer to analyze the viscoelastic properties of brain tissue. Samples were preserved in a phosphate buffered saline (PBS) in a box filled with ice during transportation and maintained at ~ 4° C before the tests. Tests were conducted at room temperature (23°C) and at body temperature (37°C) to analyze the mechanical properties of brain tissue. The measured results were clearly temperature dependent and a stiffening response was observed with decreasing temperature. Another interesting study was conducted by Zhang et al., (2011) to analyze the effects of postmortem preservation temperatures on high strain-rate material properties of brain tissues using the split Hopkinson pressure bar (SHPB). Brain samples were preserved in ice cold (group A) and in 37°C (group B) saline solution. All samples were warmed to a temperature of 37° C in a saline bath prior to testing. The stress response from brain samples preserved at 37° C was approximately 3.5 times and 2.4 times stiffer at 10% and 70% strain levels, respectively, than at ice cold.

The significant variations in stress magnitudes due to different testing temperatures (Hrapko et al., 2008) and due to different preservation temperatures (Zhang et al., 2011) may however result in inaccurate prediction of the intrinsic responses of the brain during finite element simulations. Therefore, the motivation of this study is to analyze the mechanical properties of the brain tissue at room temperature (approximately 22°C) and at body temperature (approximately 37° C) while maintaining a preservation temperature of ~ 4°C before the tests as adopted in a study by Hrapko et al., (2008). Unconfined compression tests were performed to analyze the hyperelastic behavior of the brain tissue up to 50% strain at strain rates of 30 and 50/s, using a custom designed test apparatus. There is also a need to investigate the effects of preservation temperatures, as adopted by Zhang et al., (2011), by performing unconfined compression tests to gain further insight into the behavior of brain tissue at different temperature conditions.

## 2  Materials and Methods

### 2.1  Experimental Setup

A custom made test apparatus was used in order to perform unconfined compression tests on cylindrical brain specimens (15.0 mm diameter and 6.0 mm thick) at strain rates ≤ 90/s as schematically shown in Fig. 1. A programmable servo motor controlled electronic actuator (max speed: 500 mm/s, stroke: 50 mm stroke, positioning repeatability: ± 0.02 mm, LEY 16 A, SMC Pneumatics, Ireland) was used to ensure uniform velocity during compression of brain tissue. A GSO series ± 5 N load cell (rated output of 1 mV/V nominal and safe overload of 150%, Transducer Techniques, USA) was used for the measurement of compressive force. The load cell was calibrated against known masses and a multiplication factor of 13.62 N/V was used for the conversion of voltage to load. An integrated single-supply instrumentation amplifier (AD 623 G =100, Analog Devices) with a built-in single pole low-pass filter having a cut-off frequency of 10 kHz was used. The output of the amplifier was passed through a second single pole low-pass filter with a cut-off frequency of 16 kHz. The amplified signal was analyzed through a data acquisition system with a sampling frequency of 10 kHz. The linear variable displacement transducer (LVDT) was used to measure displacement during the unconfined compression phase. The type – ACT1000A LVDT developed by RDP electronics had a sensitivity of 16 mV/mm (obtained through calibration), range ± 25 mm, linearity ± 0.25 percent of full range output.

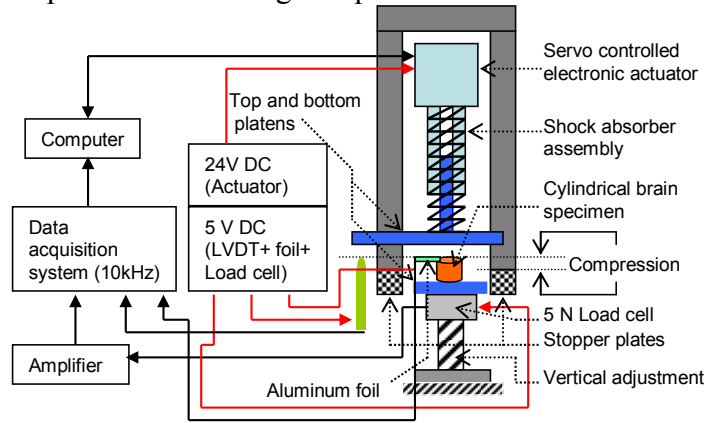

Fig. 1 – Schematic diagram of complete test apparatus for unconfined compression of brain tissue.

## 2.2 Specimen Preparation

Ten fresh porcine brains were collected from a local slaughter house and tested within 3 h postmortem. Each brain was preserved in a physiological saline solution (0.9% NaCl /154 mmol/L) at 4 to 5$^o$C during transportation. Eight specimens were prepared from each brain, in total 40 specimens were prepared from five brains. The dura and arachnoid were removed and the cerebral hemispheres were first split into right and left halves by cutting through the corpus callosum. One half of the cerebral hemisphere was cut in the coronal plane and cylindrical specimens were extracted while cutting from an anterior – posterior direction. Cylindrical specimens (15.0 ± 0.1 mm diameter and 6.1 ± 0.1 mm thick) composed of mixed white and gray matter were prepared using a circular steel die cutter. The time elapsed between harvesting the first and last specimens from each brain was 16 ~ 20 minutes for the unconfined compression tests. Not all of the specimens were excised simultaneously, rather each specimen was tested first and then another specimen was extracted from the cerebral hemisphere.

Since 8 specimens were extracted from the same brain, 4 specimens were tested at 22$^o$ C (room temperature) and remaining 4 specimens were tested at 37$^o$ C (body temperature) and same procedure was adopted for the remaining tests. Specimens were warmed to a temperature of 37$^o$ C in a saline bath for five minutes prior to testing, whereas specimens to be tested at 22$^o$ C were similarly treated with saline solution heated to 22$^o$ C. Unconfined compression tests were performed up to 50% strain. The velocity of the compression platen was adjusted to 180 and 300 mm/s, corresponding to approximate strain rates of 30 and 50/s, respectively. The attainment of uniform velocity was also confirmed during the calibration process. The top and lower platens were thoroughly lubricated with Phosphate Buffer Saline (PBS 0.9% NaCl /154 mmol/L) solution, before every test to minimize frictional effects and to ensure, as much as possible, uniform expansion in the radial direction. Each specimen was tested once and then discarded because of the highly dissipative nature of brain tissue.

## 3    Results

The cylindrical brain specimens were compressed up to 50% strain at strain rates of 30 and 50/s at room temperature (approximately 22°C) and body temperature (approximately 37°C) conditions. Preliminary force - time data obtained at each strain rate was recorded at a sampling rate of 10 kHz through a data acquisition system. The force (N) was then divided by the surface area measured in the reference configuration to determine the compressive engineering stress (Pa). Ten tests were performed at each strain rate at specific temperatures as shown in Fig. 2, in order to analyze experimental repeatability and tissue behavior. The tissue stiffness increases significantly with the increase in loading velocity, indicating the strong stress – strain rate dependency of brain tissue. The maximum compressive engineering stresses at strain rates of 30 and 50/s are 11.0 ± 2.0 kPa and 17.3 ± 2.3 kPa (mean ± SD), respectively at 22° C. Similarly, the stresses are 10.5 ± 1.29 kPa and 15.5 ± 2.2 kPa (mean ± SD) at strain rate of 30 and 50/s, respectively at 37° C. The average experimental data at different temperature conditions is also analyzed statistically using a one-way ANOVA test. Statistically, there is no significant difference (p = 0.8559) between the average stress profiles obtained at 22° C and 37° C at 30/s strain rate, as shown in Fig. 3. Similarly, there is no significant difference (p = 0.9290) due to temperature (22° C and 37° C) for the average engineering stress profiles at 50/s strain rate, as shown in Fig. 3.

We can infer stress-stretch data from Figure 2 and fit the one-term Ogden hyperelastic model (Ogden, 1972, 1997; Ogden et al., 2004) to this data for strains up to 50% with a goodness of the coefficient of determination $R^2$ of 0.99. The coefficients of the one-term Ogden strain energy density $W$ are $\mu$, the infinitesimal shear modulus (> 0), and $\alpha$, a stiffening parameter. At 37°C these are $\mu$ = 3.076kPa and 4.824kPa; $\alpha$ = 2.94 and 3.00, at 30/s and 50/s strain rates respectively. Correspondingly at 22°C and at strain rates of 30/s and 50/s, these are $\mu$ = 3.205kPa and 4.575kPa; $\alpha$ = 3.08 and 4.53.

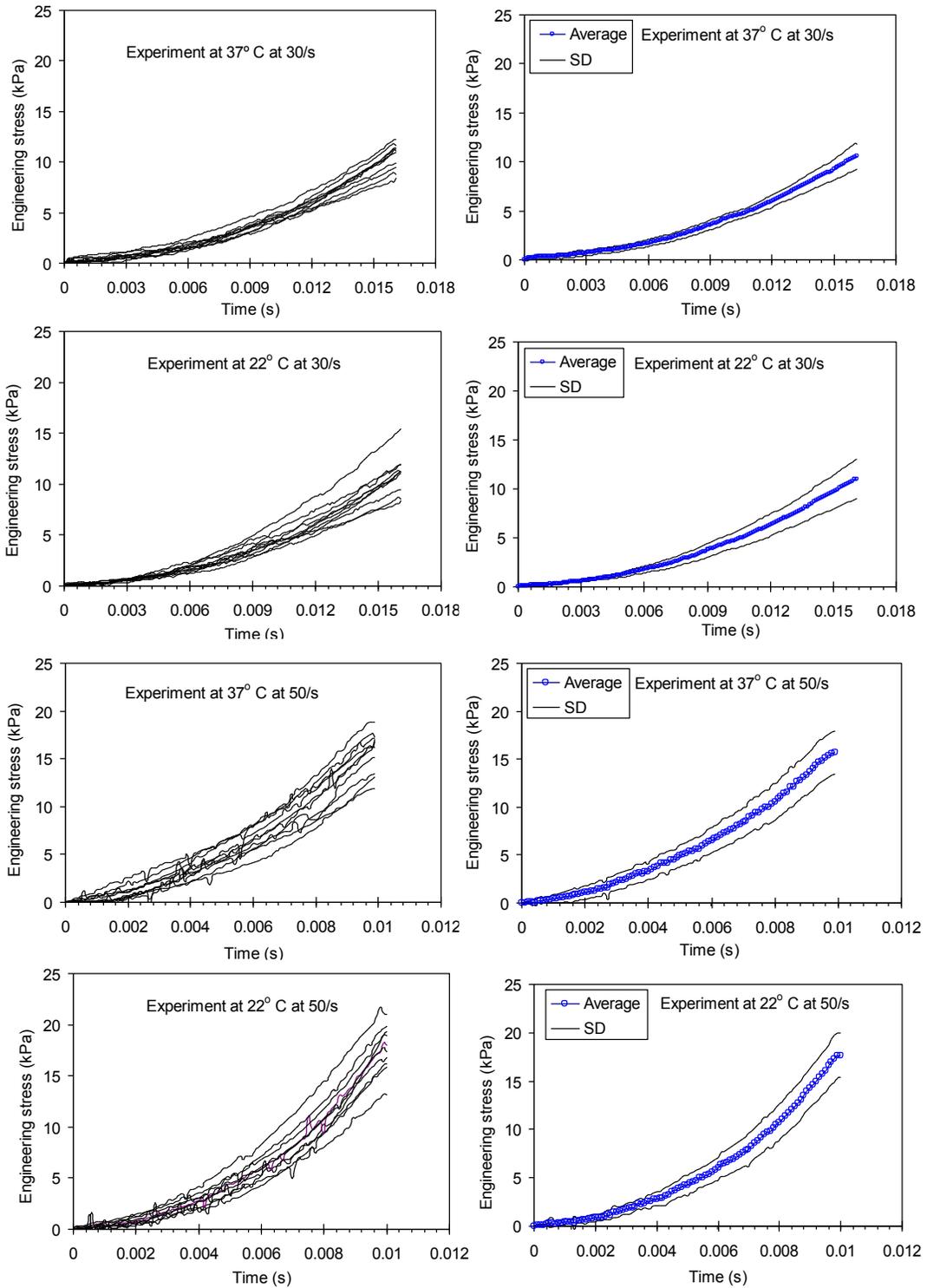

Fig. 2 – Brain tissue behavior at variable loading (30 and 50/s) and temperature conditions (22°C and 37°C)

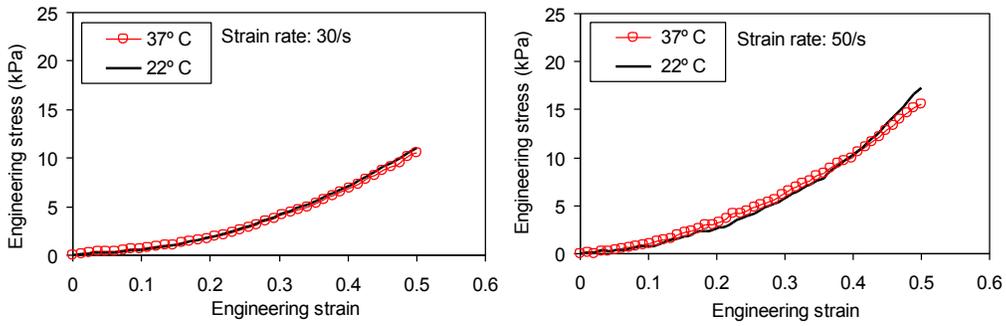
Fig. 3 – Comparison of average engineering stresses.

# 4 Discussion

Unconfined compression tests were successfully performed on porcine brain tissue up to 50% strain at room temperature (~ 22° C) and at body temperature (~ 37° C), while maintaining a constant preservation temperature of ~ 4° C. The specimens were not all excised simultaneously, rather each specimen was tested first and then another specimen was extracted from the cerebral hemisphere. This procedure was important to prevent the tissue from losing some of its stiffness and to prevent dehydration, and thus contributed towards repeatability in the experimentation. Moreover, specimens from the same brain were equally utilized for two different temperature conditions to reduce the variability because of different brains.

Similar temperature conditions were adopted by Hrapko et al., (2008) and brain tissue was tested in shear deformation in dynamic frequency sweep (DFS) and stress relaxation (SR) tests using an eccentric rotational rheometer. The measured results were found to be clearly temperature dependent with a horizontal shift factor $a_T$ between 23°C and 37°C of up to 11, whereas the vertical shift factor $b_T$ was close to one. However, in our case we have noticed a slight increase in the stiffening response of the brain tissue at variable temperature conditions, during unconfined compression tests at strain rates of 30 and 50/s. The results were analyzed statistically and we observed no significant difference (p= 0.8559 – 0.9290) in the mechanical properties of the brain tissue due to temperature variations.

A study was conducted by Huang et al., (2009) to analyze temperature effects on the viscoelastic properties of the human supraspinatus tendon using static stress - relaxation experiments at variable temperature conditions (17, 22, 27, 32, 37, 42° C). The stress profiles showed no significant difference among the six temperatures studied, implying that the viscoelastic stress response of the supraspinatus tendon was not sensitive to temperature over shorter testing durations. The instantaneous stress response of this study is similar to unconfined compression tests where time duration is significantly less.

Based on the analysis of this study, it is expected that the mechanical properties of brain tissue in unconfined compression tests are not significantly affected when tested within the approximate temperature range of 22 ~ 37° C, and particularly at short time durations at strain rates ≥ 30/s. Moreover, many research groups performed unconfined compression tests at different loading velocities within the temperature range 20 ~ 37° C (Estes and McElhaney, 1970; Miller and Chinzei, 1997; Pervin and Chen, 2009; Tamura et al., 2007).

Another interesting study was conducted by Zhang et al., (2011) specifically to analyze the effects of postmortem preservation temperatures on the properties of brain tissues using split Hopkinson pressure bar (SHPB). This method provided a continuous compression to the tissue in order to achieve 70% strain. Brain samples were preserved in ice cold and 37°C saline solutions separately in two groups. All samples were warmed to a temperature of 37° C in a saline bath and tested at the same temperature. Significant stiffening of the tissue was observed at the higher temperature (~ 37° C) instead of at the lower temperature (~ 22° C), and these findings are contrary to those of Hrapko et al., (2008). The stress response from brain samples preserved at 37° C was approximately 3.5 times and 2.4 times stiffer than ice cold at 10% and 70% strain levels, respectively. Thus there is a need to further explore the effects of variable preservation temperature conditions on the brain tissue.

The limitation of this study is the limited temperature range (22° C and 37° C) selected for the determination of mechanical properties of the brain tissue at dynamic strain rates. While a wider temperature range could be selected in future research, the range of temperatures chosen correspond to body temperature and normal ambient temperature. As discussed, there is a need to further analyze the effects of preservation temperatures also on the brain tissue. The findings of this study also may be useful for soft biological tissues, which are usually viscoelastic in nature.

## 5. Conclusions

The following results can be concluded from this study:

1 – The mechanical properties of brain tissue in unconfined compression tests are not affected significantly ($p = 0.8559 \sim 0.9290$) by variations in test temperatures ($22 \sim 37°$ C).

2 – The tissue stiffness increased significantly with the increase in loading velocity (30 to 50/s), indicating the strong stress – strain rate dependency of brain tissue.

**Acknowledgements**  The authors thank John Gahan, Tony Dennis and Pat McNally for their assistance in machining components and developing electronic circuits for the experimental setup. This work was supported for the first author by a Postgraduate Research Scholarship awarded by the Irish Research Council for Science, Engineering and Technology (IRCSET), Ireland.